\documentclass[%
 aip,
 jmp,%
 amsmath,amssymb,
 reprint,%
]{revtex4-1}

\usepackage{graphicx}
\usepackage{dcolumn}
\usepackage{bm}

\begin{document}


\title{Nuclear resonant scattering experiment with fast time response: 
new scheme for observation of $^{229\rm m}$Th radiative decay
}

\author{A. Yoshimi}
  \email{yoshimi@okayama-u.ac.jp}
  \affiliation{%
  Research Institute for Interdisciplinary Science, Division of Quantum Universe, Okayama University, 
  3-1-1 Tsushima-naka, Kita-ku, Okayama 700-8530, Japan}

\author{H. Hara}
  \affiliation{%
  Research Institute for Interdisciplinary Science, Division of Quantum Universe, Okayama University, 
  3-1-1 Tsushima-naka, Kita-ku, Okayama 700-8530, Japan}

\author{T. Hiraki}
  \affiliation{%
  Research Institute for Interdisciplinary Science, Division of Quantum Universe, Okayama University, 
  3-1-1 Tsushima-naka, Kita-ku, Okayama 700-8530, Japan}

\author{Y. Kasamatsu}
  \affiliation{%
  Department of Chemistry, Graduate School of Science, Osaka University
  1-1 Machikaneyama Toyonaka, Osaka 560-0043, Japan}

\author{S. Kitao}
  \affiliation{%
  Research Reactor Institute, Kyoto University, Kumatori-cho, Sennan-gun, Osaka 590-0494, Japan}

\author{Y. Kobayashi}
  \affiliation{%
  Research Reactor Institute, Kyoto University, Kumatori-cho, Sennan-gun, Osaka 590-0494, Japan}

\author{\\ K. Konashi}
  \affiliation{%
  Institute for Materials Research,
  International Research Center for Nuclear Materials Science, Tohoku University, 
  2145-2, Narita-cho, Oarai-machi, Higashiibaraki-gun, Ibaraki 311-1313, Japan}

\author{R. Masuda}
  \affiliation{%
  Research Reactor Institute, Kyoto University, Kumatori-cho, Sennan-gun, Osaka 590-0494, Japan}

\author{T. Masuda}
  \affiliation{%
  Research Institute for Interdisciplinary Science, Division of Quantum Universe, Okayama University, 
  3-1-1 Tsushima-naka, Kita-ku, Okayama 700-8530, Japan}

\author{Y. Miyamoto}
  \affiliation{%
  Research Institute for Interdisciplinary Science, Division of Quantum Universe, Okayama University, 
  3-1-1 Tsushima-naka, Kita-ku, Okayama 700-8530, Japan}

\author{K. Okai}
  \affiliation{%
  Graduate School of Natural Science and Technology, Okayama University,  3-1-1 Tsushima-naka, 
  Kita-ku, Okayama 700-8530, Japan}

\author{S. Okubo}
  \affiliation{%
  Graduate School of Natural Science and Technology, Okayama University,  3-1-1 Tsushima-naka, 
  Kita-ku, Okayama 700-8530, Japan}

\author{\\ R. Ozaki}
  \affiliation{%
  Graduate School of Natural Science and Technology, Okayama University,  3-1-1 Tsushima-naka, 
  Kita-ku, Okayama 700-8530, Japan}

\author{N. Sasao}
  \affiliation{%
  Research Institute for Interdisciplinary Science, Division of Quantum Universe, Okayama University, 
  3-1-1 Tsushima-naka, Kita-ku, Okayama 700-8530, Japan}

\author{O. Sato}
  \affiliation{%
  Graduate School of Natural Science and Technology, Okayama University,  3-1-1 Tsushima-naka, 
  Kita-ku, Okayama 700-8530, Japan}

\author{M. Seto}
  \affiliation{%
  Research Reactor Institute, Kyoto University, Kumatori-cho, Sennan-gun, Osaka 590-0494, Japan}

\author{T. Schumm}
  \affiliation{%
  Institute for Atomic and Subatomic Physics, TU Wien, 1020 Vienna, Austria}

\author{Y. Shigekawa}
  \affiliation{%
  Department of Chemistry, Graduate School of Science, Osaka University
  1-1 Machikaneyama Toyonaka, Osaka 560-0043, Japan}

\author{S. Stellmer}
  \affiliation{%
  Institute for Atomic and Subatomic Physics, TU Wien, 1020 Vienna, Austria}

\author{K. Suzuki}
  \affiliation{%
  Graduate School of Natural Science and Technology, Okayama University,  3-1-1 Tsushima-naka, 
  Kita-ku, Okayama 700-8530, Japan}

\author{S. Uetake}
  \affiliation{%
  Research Institute for Interdisciplinary Science, Division of Quantum Universe, Okayama University, 
  3-1-1 Tsushima-naka, Kita-ku, Okayama 700-8530, Japan}

\author{M. Watanabe}
  \affiliation{%
  Institute for Materials Research,
  International Research Center for Nuclear Materials Science, Tohoku University, 
  2145-2, Narita-cho, Oarai-machi, Higashiibaraki-gun, Ibaraki 311-1313, Japan}

\author{A. Yamaguchi}
  \affiliation{%
  RIKEN, 2-1 Hirosawa, Wako, Saitama 351-0198, Japan}

\author{Y. Yasuda}
  \affiliation{%
  Department of Chemistry, Graduate School of Science, Osaka University
  1-1 Machikaneyama Toyonaka, Osaka 560-0043, Japan}

\author{Y. Yoda}
  \affiliation{%
  Japan Synchrotron Radiation Research Institute, 1-1-1 Kouto, Sayo-cho, Sayo-gun, 
  Hyogo 679-5198, Japan }

\author{\\ K. Yoshimura}
  \affiliation{%
  Research Institute for Interdisciplinary Science, Division of Quantum Universe, Okayama University, 
  3-1-1 Tsushima-naka, Kita-ku, Okayama 700-8530, Japan}

\author{M. Yoshimura}
  \affiliation{%
  Research Institute for Interdisciplinary Science, Division of Quantum Universe, Okayama University, 
  3-1-1 Tsushima-naka, Kita-ku, Okayama 700-8530, Japan}


%




\date{\today}

\begin{abstract}
Nuclear resonant excitation of the 29.19-keV level in $^{229}$Th with high-brilliance 
synchrotron- radiation and detection of its decay signal, are
proposed with the aim of populating the extremely low-energy isomeric state of $^{229}$Th.
The proposed experiment, known as nuclear resonant scattering (NRS), 
has the merit of being free from uncertainties about the isomer level energy. 
However, it requires higher time resolution and shorter tail in the response function of the detector
than that of conventional NRS experiments because of the short lifetime of the 29.19-keV state.
We have fabricated an X-ray detector system 
which has a time resolution of 56~ps and a shorter tail function than the previously reported one.
We have demonstrated an NRS experiment with the 26.27-keV nuclear level of $^{201}$Hg for
feasibility assessment of the $^{229}$Th experiment.
The NRS signal is clearly distinct from the prompt electronic scattering signal by the implemented detector
system. 
The half-life of the 26.27-keV state of $^{201}$Hg is determined as 629~$\pm$~18~ps
which is better precision by a factor three than that reported to date.

\end{abstract}

\maketitle


\section{Introduction}

The nucleus of the thorium isotope $^{229}$Th is unusual because of the extraordinarily low energy of its first 
excited state, which is considered to be an isomeric state (hereinafter referred to as $^{229\rm m}$Th). 
This low-energy isomeric state has attracted considerable attention not
only from the perspective of nuclear and atomic physics but also because of applications to other research fields~\cite{Tkalya2015,Wense2016}. 
An important characteristic of $^{229\rm m}$Th is that it is the only excited nuclear level that is optically 
accessible among known isotopes. This optically controllable quantum state has considerable potential impact
because, unlike atomic levels, it is insensitive
to environmental disturbances 
because of the intrinsic nuclear properties of large mass,
tiny electromagnetic multipoles.  
Recently, most attention has been focused on the potential for $^{229\rm m}$Th to be used as an  
ultra-precise "nuclear clock" that could be the new frequency 
standard~\cite{Peik2003,Kazakov2014,Campbell2012}.
Proposed experiments and theoretical investigations with $^{229\rm m}$Th into possible
temporal variation of the fine-structure constant have also been reported~\cite{Berengut2009,Flambaum2006}. 
Another potential area of research is radio-chemistry, given that the isomeric decay process depends strongly on 
the chemical environment of thorium~\cite{Karpeshin2007, Kikunaga2009}. 
While the neutral  $^{229\rm m}$Th
rapidly decays through internal conversion, the charged $^{229\rm m}$Th substituted in the appropriate crystals
or confined in the ion traps can decay only through $\gamma$-emission leading to quite long life-time~\cite{Wense2016}. 
This issue is caused by the singularity that this nuclear excited energy has the same energy scale as the binding energies 
of valence electrons (7$s$ and 6$d$).

The energy and the life-time of the isomeric state had large uncertainties for many years
because of its almost degeneracy with the ground state, 
despite many experimental studies using $\gamma$-spectroscopy
of $^{233}$U $\alpha$-decay~\cite{Kroger1976,Helmer1994}.
In the past decade, however, several experimental groups have reported new results. In particular the energy of
$^{229\rm m}$Th has been measured indirectly as  $7.8\pm 0.5\ {\rm eV}$ using an X-ray spectrometer with high 
energy resolution~\cite{Beck2007,Beck2009}. In addition,  
internal conversion electrons from the isomeric state of neutral $^{229}$Th have been detected directly,  
leading to a range of possible isomeric energies of $6.3$ - $18.3$~eV~\cite{Wense2016}.
Furthermore, its internal-conversion decay half-life for the neutral atom has been determined as 
$T_{1/2}=7\pm1$~$\mu$s~\cite{Seiferle2017}. 
These results motivated us to measure the radiative 
isomeric transition to determine the isomeric energy more precisely and to determine the currently unknown 
value of the radiative life-time, which is expected to be $\sim 10^{4}$~s~\cite{Seiferle2017}.
It is important to specify these quantities as accurately as possible 
in order to pursue potential uses of $^{229\rm m}$Th in fundamental and applied sciences.

Spectroscopic measurement of the radiative emissions from an isomeric state is the best way to determine 
its energy and lifetime precisely.
Some such experimental attempts with $^{229\rm m}$Th have been reported, but these failed to detect the expected
transition signal around 7.8~eV~\cite{Jeet2015, Yamaguchi2015}. 
These experiments used ultra-violet synchrotron radiation (SR) at energies of around
7.8~eV to excite the ground state into the isomeric state in order to promote the photon emission that is associated 
with isomeric decay. The difficulty with such measurements is that nuclear excitation is affected by  
large uncertainties in both the resonant energy and the line-width (i.e., the radiative life-time of the isomeric state). 

In this paper, we describe an experimental procedure for populating the isomeric state using 29.19-keV SR to 
excite the ground state into a higher nuclear level, 
which is first suggested by Tkalya et al.~\cite{Tkalya2000}.
Then, the test experiments for feasibility assessment of such scheme with developed experimental devices 
are described. The experimental sensitivity
of this scheme with $^{229}$Th is given quantitatively.
In this excitation scheme, the isomeric state is produced through nuclear excitation whose energy and linewidth are 
both well determined, and the isomer production is independent of the values of the isomeric energy and life-time.
Furthermore, population of the isomeric state can be confirmed by detecting the photons emitted when the nuclear
transitions from the 29.19-keV excited level to the isomeric state.
 This phenomenon, which is known as nuclear resonant scattering (NRS), is a well-established method in materials 
science~\cite{Gerdau1985, Hastings1991, Seto1995, Rohlsberger2004}.
However, because of certain nuclear properties of $^{229}$Th, 
its use in NRS requires the relatively high timing resolution to have the ability to discriminate
between the NRS signal and background scattering. 
Here, we describe an experimental demonstration of such NRS with fast time response at the SPring-8
synchrotron facility in Japan using $^{201}$Hg as 
the test nucleus, which is comparable with $^{229}$Th
in terms of its excitation energy and decay properties: 
internal conversion coefficients and characteristic X-ray energies. 
The performance of this scheme is then discussed quantitatively, and the possibility of NRS experiments 
with $^{229}$Th is evaluated based on the $^{201}$Hg measurements.

\section{Nuclear resonant scattering for populating low-lying $^{229}$Th isomeric state}

\begin{figure}[htb]
\begin{center}
\includegraphics[width=0.5\textwidth]{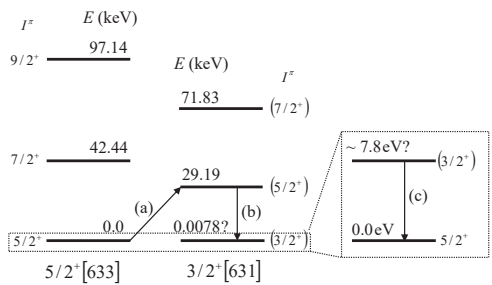}
\caption{The lower energy level structure of $^{229}$Th.}
\label{f:th-band}
\end{center}
\end{figure}

The $^{229}$Th nucleus is known to be deformed and hence to have rotational energy bands, the lower
of which are illustrated in Fig.~\ref{f:th-band}. This shows rotational bands $K^{\pi}[Nn_{z}\Lambda] = 5/2^{+}[633]$ 
and $3/2^{+}[631]$, whose band heads correspond to the ground state and the isomeric state, respectively.
Here,  the parameters $K^{\pi}$, $N$, $n_{z}$, and $\Lambda$ are the asymptotic Nilsson
quantum numbers for describing the quantum state in deformed nuclei. 
The isomeric state can be
populated from the ground state by using 29.19-keV SR to induce the M1 interband 
transition $5/2^{+} (0.0\ {\rm keV})\ \rightarrow\ 5/2^{+}(29.19\ {\rm keV})$ (transition (a) in Fig.~\ref{f:th-band}). 
This transition has fewer ambiguities in its energy and rate than those associated with 
direct excitation to the isomeric state, making it more suitable for exploring the isomeric transition.
Although this interband excitation scheme was first suggested in 2000 by Tkalya et al.~\cite{Tkalya2000} as 
a possible way to search for isomers experimentally, such experiments have never been reported. 
Tkalya et al. even calculated the expected optical activity associated with the isomeric state
(de-excitation (c) in Fig.~\ref{f:th-band}) after SR irradiation. However, they omitted an 
important issue for such a scheme to succeed, namely experimental confirmation of the interband excitation 
in advance of observing isomeric transition (c). This is important because the SR energy must be tuned to 
resonance (a) with an accuracy of $O(1\sim0.1)\ {\rm eV}$ because of the 
narrow energy width of SR. This tuning must be confirmed in advance, which can be done
by detecting the photons emitted from the $5/2^{+}(29.19\ {\rm keV})$ state
along with de-excitation (b), which is known as a measure of NRS. 
This photon detection enables us to determine the number of populated isomers if we 
assume a branching ratio from the level. Therefore, the statistics of the isomeric transition can be 
estimated reliably by detecting the NRS signal. Other advantages of this NRS-based scheme for isomer 
searching include a lower background level because of the different energy ranges for excitation and 
isomeric transition, and the ability to discriminate between genuine isomeric transitions and  
stray signals by observing changes due to detuning of the SR energy slightly.
The stray signals such as phosphorescence in the target crystal and background from other nuclear 
decays becomes problem for observing the isomeric transitions~\cite{Peik2013,Zhao2012}.

Thanks to the recent development of excellent SR facilities~\cite{Yoda2001}, 
NRS experiments with SR have become useful in material science for probing phonon spectra~\cite{Seto1995}. 
The key issue in NRS measurement is the efficient detection of the delayed NRS signal from the excited nuclear
levels following prompt scattering from the orbital electrons. 
It is thus important that the measurement system has a high temporal resolution to discriminate between
these two processes. 
Furthermore, 
a slow tail component appearing in the detector time response
in addition to a Gaussian function tends to affect such discrimination.
Because the cross section of the prompt scattering
is generally 6-7 orders of magnitude greater than that of the nuclear excitation, 
this tail component must be small especially for measurement of the short-life nuclear levels.
The excited nuclear states used for NRS thus far have been limited to levels whose half-life are longer 
than $\sim$0.6~ns. However, the half-life of the $5/2^{+}(29.19\ {\rm keV})$ state of $^{229}$Th has not 
been measured 
because of the small branching ratio of alpha decay of $^{233}$U into this state. 
Nevertheless, it can be estimated using either reported $g$-factor values for 
the rotational band (i.e., $|g_{K}-g_{R}|=0.59(14)$ ~\cite{Kroger1976}) or 
the relevant M1 reduced transition probability (i.e., $B({\rm M1})=0.048(5)\ \mu_{\rm N}^{2}$ ~\cite{Barci2003}).
With an internal conversion coefficient $\alpha=225$ (assuming a value of M1/E2 mixing ratio of 
$\delta=0.145$ ~\cite{Barci2003}) for this transition~\cite{ICC}, the half-life of the state is estimated to be
\begin{eqnarray}
T_{1/2} &=& \frac{\ln{2}}{\Gamma_{\gamma, \rm M1}\cdot(\alpha +1)}\cdot\frac{12}{13} \nonumber \\
&=& \frac{12/13\times \ln{2}}{1.76\times 10^{13}E^{3}_{\gamma}B({\rm M1})\cdot (\alpha+1)} 
\simeq 0.15\ {\rm ns}.
\end{eqnarray}
Here, $\Gamma_{\gamma, \rm M1}$ is the M1 radiative transition rate from the 29.19-keV level to the
isomeric state, $E_{\gamma}$ is the transition energy in units of megaelectron volts~\cite{BohrMot}. 
The factor $12/13$ is introduced in the calculation on the assumption of a branching ratio of 
$12/13$~\cite{Beck2007} from this state to the isomeric state.
(It is noted that the estimated branching ratios in the several reports show spread~\cite{Tkalya2015}.)  
This estimated half-life necessitates NRS with nuclear levels whose half-life are shorter than 
any ever used in previous experiments if the proposed experiment is to be conducted successfully. 

Also important are the reaction rates of the NRS and prompt scattering, 
where the values are determined
by the incoming photon flux density, number of atoms, and each reaction cross section.
The linewidth of the excitation from the ground 
state to the $5/2^{+}(29.19\ {\rm keV})$ state is estimated as follows based on the 
above estimated half-life, internal conversion coefficient, and branching ratio of this state;  
\begin{equation}\label{eq:line-width}
\Gamma_{\gamma} = \frac{1}{12}\cdot\frac{1}{1+\alpha}\cdot\frac{\hbar \ln{2}}{T_{1/2}} \simeq 1.7\ {\rm neV}.
\end{equation}
Here, the internal conversion coefficient $\alpha$ for the relevant interband 
transition $5/2^{+}(0.0\ {\rm keV})\leftrightarrow 5/2^{+}(29.19\ {\rm keV})$ is uncertain because of
unknown M1/E2 mixing. We here use the calculated value $\alpha=149$~\cite{Barci2003}.
The effective NRS cross section at X-ray energies
tuned to resonance is described as
\begin{equation}\label{eq:nrs-equ}
\sigma_{\rm Th\mathchar` NRS} = \frac{\lambda^{2}}{2\pi}\frac{2I_{\rm e}+1}{2I_{\rm g}+1}
\frac{\Gamma_{\gamma}}{\Gamma_{0}} \cdot\frac{\Gamma_{0}}{\Gamma_{\rm Xray}},
\end{equation}
which includes the effect of the X-ray bandwidth. Conventionally, this allows the NRS cross section
to be compared directly to the photoelectric scattering cross section. 
Here, $\lambda$ is the wavelength of the incoming X-rays, and $I_{\rm g}$ and $I_{\rm e}$ are the nuclear spins
in the ground state and excited state, respectively. The term $\Gamma_{0}$ is the natural linewidth, 
which is related to the (1/e) decay
time from the excited state to the ground state. The ratio $\Gamma_{\gamma}/\Gamma_{0}$ is therefore 
the probability that the excited nucleus decays through $\gamma$-ray emission.  
Using the internal conversion coefficient, this ratio can be written as 
$\Gamma_{\gamma}/\Gamma_{0} = 1/(1+\alpha)$. The factor $\Gamma_{0}/\Gamma_{\rm Xray}$ is 
introduced as the excitation efficiency of the incoming X-rays, the width of whose bandwidth 
$\Gamma_{\rm Xray}$ is typically many orders of magnitude broader than the natural width $\Gamma_{0}$.
The NRS cross section of the relevant $^{229}$Th excitation is thus 1.3~mb
(1~b = $10^{-28}$~m$^{2}$) when taking the typical 
singly monochromatized bandwidth $10^{-4}$ ($\Gamma_{\rm Xray}= 4$~eV) of SR. 
In contrast, the prompt scattering has larger cross sections than that of NRS, where the photoelectric 
absorption is a main process while the contribution of the Compton scattering and the Rayleigh scattering 
are negligibly small at the X-ray energies around 29~keV.
The cross section of the photoelectric absorption that produces prompt fluorescence in Thorium is 
$\sigma_{\rm photoe}=15.4$~kb at a photon energy of 29.19~keV~\cite{xcom}.
Therefore, the NRS signal has to be observed separately from the prompt signal given that the cross section 
of the former is seven orders of magnitude smaller than that of the latter.

The essential choice then becomes that of a detector whose time response has both a high temporal resolution
and a short tail. Conventional NRS measurement uses counting gates after
the SR excitation pulse (the full width at half maximum (FWHM) of which is typically 30-40~ps) to remove huge 
prompt background events. 
The prompt scattering itself from atomic electrons takes place on a subpicosecond time scale in the case of 
29.19-keV radiation with a thorium target, whereas the NRS time scale is the lifetime of the nuclear levels.
An Si avalanche photo diode (APD) of diameter of 1--3~mm and depletion-layer depth of 
10--30~$\mu$m is often used in NRS experiments; the typical time resolution of such an APD is 
100--200~ps~\cite{Kishimoto1994, Baron2006}.
The counting gates start typically 2--5~ns after the synchrotron pulse in order to separate the NRS and prompt
events~\cite{Seto2000, Ishikawa2005, Bessas2015}. However, 
the detector system for NRS measurement of $^{229}$Th requires a higher time resolution because of 
the especially short nuclear life-time.
In order to demonstrate the feasibility of the proposed NRS experiment with such a short nuclear life-time,
we show that it is possible to have clearly separated NRS and prompt events within a short elapsed time.
We do this by collecting all the prompt and NRS scattering signals, without reducing their statistics,
by using APD detectors that are faster than those used previously.

\section{Nuclear resonant scattering experiment using $^{201}$H{g}}

The half-life of the second 26.27-keV-excited level of $^{201}$Hg is reported to be 
$T_{1/2}=0.630(50)$~ns~\cite{Schuler1983}. This is the shortest half-life of all the nuclear levels
measured in NRS experiments~\cite{Ishikawa2005}, and therefore this level is the best target for this 
demonstration in view of its short lifetime and comparable excitation energy.
Its atomic number $Z=80$ and large internal conversion coefficient $\alpha = 71.6$~\cite{nndc} 
are also comparable to those of the relevant $^{229}$Th level. The parameters related to the nuclear resonance 
of $^{201}$Hg and
$^{229}$Th are summarized in Table~\ref{tab:nrs-param} for comparison.
The excitation linewidths are calculated using Eq.~(\ref{eq:line-width}) with a unit branching ratio for $^{201}$Hg
because decay to the first excited 1.565-keV state ($1/2^{-}$) is negligible.
\begin{table}[htb] 
\caption{Parameters related to NRS experiments with $^{229}$Th and $^{201}$Hg. Internal conversion
coefficients for sub-shell components are from \cite{ICC}. Values of fluorescent yields 
are from \cite{Ferreira1971,Shatendra1984} for thorium and from \cite{Hubbell1994,Bambynek1972} 
for mercury. Photoelectric absorption cross sections are from \cite{xcom}.}
\begin{center}
\begin{tabular}{c|c|c}\hline\hline
Nucleus & $^{229}$Th & $^{201}$Hg  \\
\hline\hline
Excitation energy & 29.1927(5) keV & 26.272(25) keV \\
\hline
$I^{\pi}_{\rm g} \rightarrow I^{\pi}_{\rm e}$ & $5/2^{+} \rightarrow 5/2^{+}$ & $3/2^{-} \rightarrow 5/2^{-}$ \\ 
\hline
Excitation width & 1.7 neV & 10.0 neV \\
\hline
$T_{1/2}$ in excited state & $\simeq 0.15$\ ns & 0.630(50) ns \\
\hline
\raisebox{0.5em}{\shortstack{Internal conversion \\ coefficient}} & 
\shortstack{$\alpha_{\rm total}=225$ \\ $\alpha_{L}=167.9$ \\ 
$\alpha_{M}=42.5$} & 
\shortstack{$\alpha_{\rm total}=71.6$ \\ $\alpha_{L}=55.5$ \\ $\alpha_{M}=13.0$}  \\
\hline
\raisebox{0.5em}{Fluorescence yield} & \shortstack{$\omega_{L}=0.46(3)$ \\ $\omega_{M}=0.044(4)$}  & 
\shortstack{$\omega_{L}=0.35(1)$ \\ $\omega_{M}\simeq 0.026$} \\
\hline
\shortstack{Photoelectric absorption \\ cross section \\ at respective NRS energy} & 
\raisebox{1em}{15.4\ {\rm kb}} & \raisebox{1em}{12.8\ {\rm kb}} \\
\hline\hline
\end{tabular}
\end{center}
\label{tab:nrs-param}
\end{table}
The internal conversion coefficients for sub-shell components and the fluorescent yields for each sub-shell,
which are important parameters for investigating NRS count rates, are also summarized in the table. 

\begin{figure}[htb]
\begin{center}
\includegraphics[width=0.5\textwidth]{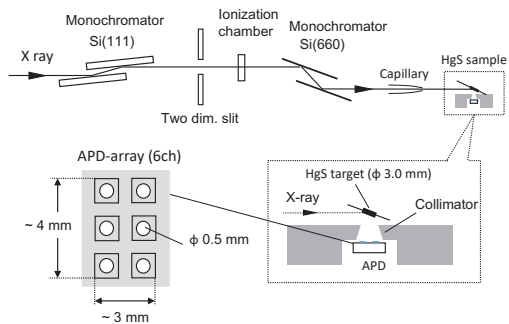}
\caption{Experimental setup for NRS of $^{201}$Hg nucleus.}
\label{f:setup1}
\end{center}
\end{figure}

The present experiments were performed at the BL09XU beam line of SPring-8. The electron beam current
in the storage ring was 100~mA, and the ring was operated in a 203-bunch mode with a 23.6~ns interval.
The measured bunch width was 35~ps at FWHM.
The X-rays from the undulator were doubly monochromatized by Si(111) and Si(660) monochromators
as shown in Fig.~\ref{f:setup1}. 
The X-rays reflected by the first Si(111) monochromator has a maximum intensity of
$4\times 10^{13}$~photons/s and a bandwidth of 3.4~eV (FWHM) at an energy of 
26.27~keV. The angle of the Si(111) was set so as to 
obtain the 26.27-keV X-ray.
The second Si(660) monochromator was used to reduce the bandwidth further.
The incident X-ray beam was collimated to 0.8~mm~$\times$~0.8~mm by a two-dimensional slit,
and focused to a spot size of 0.3~mm~$\times$~0.2~mm at the target position 
by a tapered glass capillary (HORIBA, 2014SP13).
The photon flux was monitored nondestructively using a small ion chamber, whose current was calibrated using a 
PIN photo diode (not shown in the figure).
The photon fluxes measured at the target position with and without the capillary were
$3.2\times10^{11}$~photons/s and $3.7\times 10^{11}$~photons/s, respectively.
An HgS powder with a 13.18 \% natural abundance  of $^{201}$Hg was used as the target material, which was located 
downstream of the beam line. 
A 4.8 mg quantity of HgS was pelletized to a disk of 3.0~mm diameter and 0.08~mm thickness. The number densities of 
$^{\rm nat}$Hg and $^{201}$Hg are $2.1\times 10^{22}$~cm$^{-3}$ and $2.8\times 10^{21}$~cm$^{-3}$, respectively.

Because of the large internal conversion coefficient, the
NRS signal consists mainly of characteristic L X-rays following internal conversion of the excited nuclear 
state. These fluorescences are distributed in an energy range of 10--15~keV: 
$L_{\alpha}=9.9$~keV, $L_{\beta}=11.8$~keV, $L_{\gamma}=13.6$~keV. These signals coincide completely with 
the prompt signal in relation to energy because the incident 26.27-keV X-rays induce photoelectric processes 
in the same electron shell. As summarized in Table~\ref{tab:nrs-param}, the internal conversion of M-shell
electrons also occurs with a coefficient of $\alpha_{M}=13$. However, the fluorescence following M-shell conversion
is not observable because of the relatively small fluorescence yield of $\omega_{M}\simeq 0.026$. 
An Si-APD  (Hamamatsu Photonics, S12053-05) with a small diameter of 0.5~mm and a thin depletion layer of 
10~$\mu$m which is estimated from the data sheet, was used to detect this fluorescence. 
Such small and thin APD detectors tend to have relatively low detection efficiency, but fast time responses.
Six Si-APD chips were mounted on a fabricated substrate to expand the sensitive area,
as shown in Fig.~\ref{f:setup1}. These were placed 3.5~mm away from the target sample, which was tilted 
at an angle of $25^{\circ}$ to the 
beam direction. The solid angle of the APD-array system was $\Omega \simeq 80$~mstr. 
A cone-shaped brass collimator with diameters of 1.2~mm and 3.0~mm, was placed between the target 
and the detector to reduce background X-ray scattering from different angles.
Each APD output was amplified by a fast amplifier and was processed to a digital signal by a fast constant-fraction 
discriminator (CFD). 
Each event was labeled with its arrival time by a multi-stop time-to-digital converter (TDC) 
(FAST ComTec, MCS6) with a resolution of 50~ps, 
and was stored in a PC. This data acquisition was processed in parallel from the six APD chips to PCs.   
The analogue pulse height of the APD signal was taken simultaneously in the TDC by using 
a fabricated analogue-to-time converter (ATC). Two-dimensional analysis based on time and pulse height can 
reduce the background signal. This data-taking system can accept event rates up to 
$\sim$1~MHz for each APD channel. 
Further details about the detector and data-acquisition system are given in \cite{Masuda2017}.

The rate of fluorescence detection associated with an NRS event is given by
\begin{equation}
\label{eq:NRS-cross-section}
R_{\rm Hg\mathchar` NRS} = N_{\rm 201Hg}\sigma_{\rm Hg\mathchar` NRS} \Phi_{\rm Xray} \epsilon_{\rm LXray}
\epsilon_{\rm eff},
\end{equation} 
where $\Phi_{\rm Xray}$ is the average photon flux density on the target, which was measured as 
$\Phi_{\rm Xray}\simeq 1.4 \times 10^{12}$~photon/s/mm$^{2}$ within an area of 0.22~mm$^{2}$.
The number of $^{201}$Hg atoms irradiated by X-rays was estimated to have been 
$N_{\rm 201Hg} \simeq (3.7\times 10^{17})\times 0.1318$ by considering an X-ray
attenuation length of 0.038~mm in the Hg target. 
The emission probability $\epsilon_{\rm LXray}$ of L-shell fluorescence is calculated to be
$\epsilon_{\rm LXray}=(\alpha_{L}/\alpha_{\rm total})\omega_{L}=0.268$. 
The efficiency parameter $\epsilon_{\rm eff}$ includes the APD detection efficiency, the solid-angle efficiency, 
transmission ratio of fluorescence in the brass collimator, and fluorescence attenuation in the target. 
This parameter is estimated roughly as $\epsilon_{\rm eff}\sim 1\times 10^{-5}$.
The NRS cross section $\sigma_{\rm Hg\mathchar` NRS}$ is estimated as 359~mb with Eq.~(\ref{eq:nrs-equ}) using 
the parameter values given in Table~\ref{tab:nrs-param}. The linewidth $\Gamma_{\rm Xray}$ 
of the incident X-rays is determined to be 148 meV by measuring the energy spectrum of the NRS signal,
as described in the following section.

\begin{figure}[htb]
\begin{center}
\includegraphics[width=0.45\textwidth]{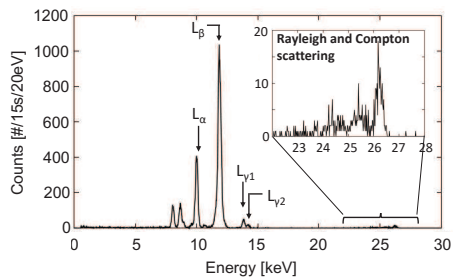}
\caption{Fluorescence energy spectrum from the HgS target with 26.27-keV X-ray irradiation. The small peaks 
around 8~keV are characteristic X-ray of Cu scattered from the brass around the target.}
\label{f:Hg-SDD}
\end{center}
\end{figure}

The prompt fluorescence signal follows the photoelectric absorption in all Hg isotopes. The total 
photoelectric cross section is $\sigma_{\rm photoe}=12.8$~kb at an energy of 26.27~keV 
(Table~\ref{tab:nrs-param}).
Therefore, the prompt-fluorescence detection rate is estimated by replacing 
$N_{\rm 201Hg}$ and $\sigma_{\rm Hg\mathchar` NRS}$ by $N_{\rm Hg}=(100/13.18)N_{\rm 201Hg}$ and
$\sigma_{\rm photoe}$, respectively, in Eq.~(\ref{eq:NRS-cross-section}). 
The cross sections for the electron shells at this energy are obtained as 
$\sigma_{L}=10.14$~kb and $\sigma_{M}=2.48$~kb by interpolating from calculated data sheet~\cite{Scofield1973}. 
The L-X-ray emission probability is thus estimated as 
$\epsilon_{\rm LXray}=(\sigma_{L}/\sigma_{\rm photoe})\omega_{L}=0.279$. The L-shell fluorescence probability
for NRS and prompt events is then almost same within 0.27--0.28. Therefore,
the relative NRS/prompt signal intensity is estimated extremely accurately, it being unaffected by 
the uncertainty in $\epsilon_{\rm eff}$. 
A typical energy spectrum of the prompt event is shown in Fig.~\ref{f:Hg-SDD}.
It was measured using an Si drift detector (SDD) at the incident X-ray energy 
of 26.27~keV, to confirm the origin of the measured fluorescence.
This indicates that the observed fluorescence consisted almost entirely of characteristic X-ray
of the L-shell. The measured amounts of Rayleigh and Compton scatterings of the incoming X-ray were no more 
than $2\%$ for all the measured fluorescence signals.

\section{Results and discussion}

\subsection{Results of  NRS experiments with $^{201}$Hg}

\begin{figure}[hbt]
\begin{center}
\includegraphics[width=0.45\textwidth]{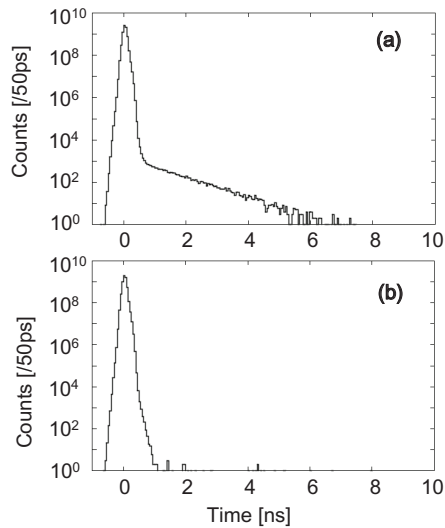}
\caption{Typical time spectrum of fluorescence signal from HgS target with incident X-ray energy 
tuned to nuclear excitation of $^{201}$Hg. Integration time was 2,000~s for each spectrum.}
\label{f:time-spectrum}
\end{center}
\end{figure}

A typical time spectrum measured when the incident energy was tuned to the  
$3/2^{-}(0 {\rm keV}) \rightarrow 5/2^{-}(26.27 {\rm keV})$ transition of the $^{201}$Hg nucleus
is shown in Fig.~\ref{f:time-spectrum} (a).
This spectrum contains both the prompt peak and the delayed NRS signal, the amplitude of the latter
being six orders of magnitude smaller than the amplitude of the former.
The time resolution of the whole system was
evaluated by $\chi$-squares fitting a Gaussian function to the prompt peak.
The standard-deviation width was found to be $\sigma = 56(2)$~ps. 
The intrinsic time resolution of the detector system was thus estimated to be $\sigma_{\rm detector}=54$~ps 
by taking account of the SR pulse width of $\sigma_{\rm SR} = 15$~ps. 
The tail behavior of the prompt peak was evaluated from the time spectrum with the incident energy being far
from resonance with the transition energy (Fig.~\ref{f:time-spectrum} (b)).
Here, the off-resonance time spectrum was taken with the incident energy detuned by 0.44~eV from the 
resonance center; it has a very short tail of less than 1~ns at $10^{-9}$. 
This high time resolution and short tail compared to the previous measurements were successfully realized
by fabricating the detector and data-acquisition system specifically for the required timing resolution.
This short tail of the detector time response enabled us to separate the pure NRS component from the prompt 
spectrum in a short elapsed time.
The pure NRS spectrum was extracted in a time window of 0.9--5.2~ns and the half-life
was then determined by $\chi$-squares fitting a single exponential function to the NRS spectrum.
We found that the reduced-$\chi^{2}$ ranged within 0.9--1.1 in 
the different fitting regions; the minimum range of 1.8--3.0~ns 
and the maximum range of 0.9--5.2~ns.  
The statistical and the systematic errors in the fitting procedure were obtained 14~ps and 10~ps, respectively. 
A time jitter of 6~ps in the standard clock at SPring-8~\cite{Kawashima2008}, that was
utilized in our data-acquisition system, produces additional systematic effect. 
The half-life was then determined with the total uncertainty which is the quadratic sum of the statistical uncertainty and
the systematic uncertainties, as
\begin{equation}
T_{1/2}(26.27\ {\rm keV}) = 629\ \pm 18\ {\rm ps}. 
\end{equation}
This precision is better than the previous one that was determined from a Coulomb-excitation 
experiment~\cite{Schuler1983} by a factor of three, and the value is consistent within the error.

\begin{figure}[bth]
\begin{center}
\includegraphics[width=0.45\textwidth]{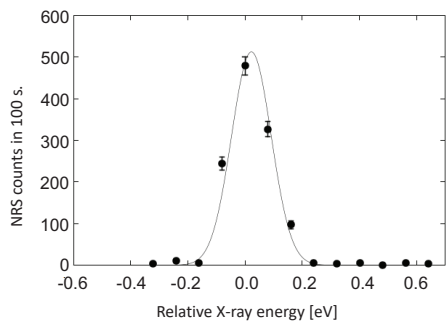}
\caption{NRS count as a function of incident X-ray energy controlled by scanning the Si monochromators.
The width was found to be $148 \pm 22$~meV (FWHM).}
\label{f:energy-spectrum}
\end{center}
\end{figure}

The resonance spectrum was obtained by plotting the delayed NRS counts in a time window of
1.0--5.0~ns along the incident X-ray energy, as shown in Fig.~\ref{f:energy-spectrum}. The X-ray energy
was scanned by controlling the angle of the Si(660) monochromator in steps of 0.26~s, which corresponds to 
an energy step of 83~meV.
The width of this spectrum is dominated by the X-ray bandwidth (as determined by the double monochromator 
system) because this bandwidth is larger than the width of the nuclear level by many order of magnitude. 
The energy resolution of the monochromator system was determined as 
$\Gamma_{\rm Xray} = 148 \pm 22$~meV (FWHM) by fitting a Gaussian function to each of the observed
spectra and taking the average.

Measuring the whole time spectrum including both the prompt and delayed NRS components is effective for 
evaluating the measurement system for application to other nuclear targets such as $^{229}$Th.
The ratio of the prompt count rate to the NRS one is expressed simply as 
$R_{\rm NRS}/R_{\rm prompt} = (N_{\rm 201Hg}/N_{\rm Hg})\times(\sigma_{\rm NRS}/\sigma_{\rm photoe})$, 
as described previously. 
From the above determined bandwidth of the incident X-rays, the effective NRS cross section 
(Eq.~\ref{eq:NRS-cross-section}) for $^{201}$Hg is estimated as $\sigma_{\rm Hg\mathchar` NRS} = 359$~mb. This 
gives the ratio $R_{\rm NRS}/R_{\rm prompt} = 3.6\times10^{-6}$, which is found to be consistent with the measured 
count rates of $R_{\rm NRS}=14$~Hz and $R_{\rm prompt}=3.6$~MHz in the spectrum shown in  
Fig.~\ref{f:time-spectrum} (a). 
Both count rates are obtained consistently if the detector efficiency including the sensor area coverage and the 
fluorescence transmission through the collimator is $\epsilon_{\rm eff} = 2.1\times 10^{-5}$ based on the measured 
X-ray flux density and estimated numbers of irradiated atoms, $N_{\rm Hg}=3.5\times 10^{17}$ and 
$N_{\rm 201Hg} = 4.6\times10^{16}$. The photon flux, which was monitored continuously in these measurements, was 
stable to within $1.5\%$ (peak- to- peak) and was sufficiently reliable for the above estimation.

\subsection{Discussion of NRS measurements with $^{229}$Th}

The NRS count rate for $^{229}$Th nuclei can be estimated based on the NRS detection rate of 14~Hz
in the present $^{201}$Hg experiments, neglecting any uncertainty in detector efficiency.
The test experiments for the $^{229}$Th-NRS measurement have been performed using the purified 
$^{229}$Th target~\cite{Masuda2017}. This target was prepared as deposited 4.4~$\mu$g-$^{229}$Th(OH)$_{4}$ 
on a polypropylene sheet.
With the same amount of $^{229}$Th target  (in which the number of $^{229}$Th nuclei was $7.9\times 10^{15}$) 
in the $\phi$~0.3 mm spot and 29-keV X-rays monochromatized by the same setup used in the present 
experiment focused onto this size of target, the NRS detection rate becomes
\begin{eqnarray}
R_{\rm Th\mathchar` NRS} &=& 14\ {\rm Hz}\times 
\frac{\sigma_{\rm Th\mathchar` NRS}}{\sigma_{\rm Hg\mathchar` NRS}}\cdot
\frac{N_{\rm 229Th}}{N_{\rm 201Hg}}\cdot \frac{\Phi_{\rm Xray\mathchar` Th}}{\Phi_{\rm Xray\mathchar` Hg}} 
\nonumber \\
& \simeq & 0.8\ {\rm Hz}.
\end{eqnarray}
Here, the effective NRS cross section for $^{229}$Th is 
$\sigma_{\rm Th\mathchar` NRS} = 33.3$~mb with the same 
X-ray bandwidth as that in the present experiments. The photon flux density on the Th target is assumed 
$\Phi_{\rm Xray\mathchar` Th} = 3.7\times 10^{11}\ {\rm photons/s}/(\pi\cdot0.15^{2}\ {\rm mm^{2}}) = 
5.2\times 10^{12}$~photons/s/mm$^{2}$.
An integrated thorium target can be prepared by the same deposition method as~\cite{Masuda2017}, or
electrodeposition scheme~\cite{Haba2006}. Both types of thorium target have been found to be sufficiently 
stable for beam-irradiating experiments.
The detector solid angle can be improved by a factor of two by introducing more APD chips with higher integration,
leading to an expected NRS rate of $\simeq 1.5$~Hz.
The count rate of the prompt process in a $^{229}$Th experiment is thus estimated to be 
$1.5\ {\rm Hz} \times (15.4~{\rm kb}/33.3~{\rm mb}) \simeq 0.7$~MHz.
With these estimated statistics,
the NRS spectrum and the prompt tail are expected to be observed with counts of 
$\sim$10 and $\sim$1, respectively, at $t=1$~ns in an integration time of 2,000~s. 
For  $^{229}$Th, the NRS and prompt events would therefore be observed separately
in a few thousand seconds using the present experimental scheme. 
The constant background coming from the radioactivity of $^{229}$Th and its 
daughter elements could be reduced by selecting the observed fluorescence energy in the analysis~\cite{Masuda2017}.
The isomeric transition of $^{229}$Th could be detected by switching from a thorium target to 
a VUV-transparent Th-doped crystal such as MgF$_{2}$ or CaF$_{2}$~\cite{Stellmer2015} after confirming the 
detection of the NRS signal. This NRS detection is direct evidence that the X-ray energy is tuned precisely to 
nuclear resonance and that the isomeric state of $^{229}$Th is populated.

\section{Conclusion}

The 29.19-keV nuclear excitation of $^{229}$Th with high-brilliance SR and the
measurement of its decay signal, known as NRS, were investigated as a mean of
populating an extraordinarily low-energy isomeric state. 
The half-life of the 29.19-keV excited state was estimated to be as short as 0.15~ns, which is a timescale on
which it is difficult to perform conventional NRS measurements. Consequently,
NRS measurements with a better time response were proposed for measuring the isomeric transition.  
This was achieved by fabricating a fast 
detector system. The 26.27-keV nuclear excitation of $^{201}$Hg was used for NRS measurements to assess 
the performance of all the devices and to test the feasibility of applying this scheme to NRS with $^{229}$Th. 
The measured time resolution of the entire system was 56~ps in standard deviation, 
and the time response had a relatively short tail of 1~ns at $10^{-9}$. Both NRS and
prompt electronic scattering were observed with a high count rate of 3.6~MHz, and thus the half-life of the 
26.27-keV excited state was determined precisely as $T_{1/2}(26.27\ {\rm keV}) = 629 \pm 18$~ps 
thanks to the high timing resolution. 
The NRS count rate of $^{229}$Th was estimated as $\sim1.5$~Hz, based on the statistics of the present 
experiment and assuming target preparation with a well-established method. The present scheme was found
to be feasible for populating the isomeric state of the $^{229}$Th nucleus.

\begin{acknowledgments}
The synchrotron radiation experiments were performed at the BL09XU of SPring-8 with the approval of the 
Japan Synchrotron Radiation Research Institute (JASRI)  (Proposal No. 2014A1334, 2014B1254, 2015B1380, 
2016A1420, and 2016B1232).
This work was supported by JSPS KAKENHI Grant Numbers 15H03661, 17K14291, 24221005, 
Technology Pioneering Projects in RIKEN, and MATSUO FOUNDATION. 
S. Stellmer and T. Schumm acknowledge support by the EU-FET-Open project 664732 NuClock.
\end{acknowledgments}


\nocite{*}


\end{document}